\documentclass[pra,aps,twocolumn,floatfix,showpacs]{revtex4-1}
\usepackage{graphicx,amsmath,amssymb,times}

\begin{document}
\title{Isothermal sweep theorems for ultra-cold quantum gases in a canonical ensemble}
\author{M. Iskin}
\affiliation{Department of Physics, Ko\c c University, Rumelifeneri Yolu, 34450 Sariyer, Istanbul, Turkey.}
\date{\today}

\begin{abstract}

After deriving the isothermal Hellmann-Feynman theorem (IHFT) that is suitable 
for mixed states in a canonical ensemble, we use this theorem to obtain 
the isothermal magnetic-field sweep theorems for the free, average 
and trapping energies, and for the entropy, specific heat, pressure and 
atomic compressibility of strongly-correlated ultra-cold quantum gases. 
In particular, we apply the sweep theorems to two-component Fermi gases 
in the weakly-interacting BCS and BEC limits, showing that the 
temperature dependence of the contact parameter can be determined by 
the variation of either the entropy or specific heat with respect to 
the scattering length. We also use the IHFT to obtain the Virial 
theorem in a canonical ensemble, and discuss its implications for 
quantum gases.

\end{abstract}

\maketitle

\section{Introduction}
\label{sec:introduction}

Dilute samples of ultra-cold quantum gases are well-described by zero range 
interactions, a consequence of which is the existence of some universal 
relations that govern the behavior of these systems~\cite{review}. 
The first examples of such relations were originally derived by solving the 
many-body Schr\"odinger equation~\cite{tan1, tan2, tan3}, and they relate energy and pressure of
the system to a single parameter that comes from the short-range physics.
What makes these relations remarkable is that they hold for any finite-energy 
state of the system, it does not matter whether the system is few- or many-body, 
superfluid or normal, weakly or strongly interacting, in equilibrium or 
nonequilibrium, at zero or finite temperature. 
These universal relations have more recently been rederived using many other 
approaches including the quantum field theoretical 
techniques~\cite{review, braaten1, braaten2, zhang, combescot, werner2},
and have also been verified via numerical few-body calculations~\cite{blume}. 

There is one common element in these universal relations: they all 
involve a single parameter called the contact. In the case of two-component 
Fermi gases, the contact parameter is given by the coefficient of the $1/k^4$ tail of 
the momentum distribution, and therefore, it measures the number of fermion 
pairs with small separations. It has been found that the contact parameter 
also appears in many other short-range (high-momentum) or short-time (high-frequency) 
properties of the system~\cite{baym, haussmann, hu1, son1,taylor, schneider}.
This parameter has recently been measured in an ultracold $^{40}$K gas
via the measurements of the high-momentum tail of the momentum distribution 
and of the high-frequency tail of the radio-frequency signal~\cite{stewart}, 
and also measured in an ultracold $^6$Li gas via the measurements of the static
structure factor~\cite{hu2}. The measured temperature and scattering length 
dependence of the contact parameter compare well with the theoretical 
predictions~\cite{yu, palestini}.

In particular, the so-called adibatic sweep theorem for the energy~\cite{tan2}, 
which has recently been experimentally verified~\cite{stewart}, 
relates the contact to the change in total energy of the system 
when the atom-atom scattering length ($a$) is changed adiabatically, i.e. 
with zero heat transfer.
In atomic systems, the value of $a$ can be tuned at will from 
very large and negative to very large and positive values, thanks to 
the presence of magnetically induced Feshbach resonances, and an 
adiabatic (constant entropy) sweep is accomplished by changing the 
bias magnetic field over a time scale longer than the relaxation time~\cite{stewart}.

Motivated by these earlier works, here we generalize a number of these 
universal relations to finite temperatures, and
the rest of the manuscript is organized as follows. In Sec.~\ref{sec:hft},
we derive the isothermal Hellmann-Feynman theorem (IHFT) that is suitable for mixed 
states in a canonical ensemble. Then, we use this theorem in 
Sec.~\ref{sec:apps} to obtain the isothermal (constant temperature) sweep 
theorems (IST) for the free, average and trapping energies, and for the entropy, 
specific heat, pressure and atomic compressibility of strongly-correlated ultra-cold 
Fermi gases. We also use the IHFT to derive the Virial theorem in a 
canonical ensemble in Sec.~\ref{sec:virial}.
A brief summary of our conclusions is given in Sec.~\ref{sec:conclusions}.

\section{Hellmann-Feynman theorems}
\label{sec:hft}

For our purpose, we need to derive the Hellmann-Feynman theorem (HFT) 
suitable for a canonical ensemble of a fixed number $N$ of identical particles 
that are in thermal equilibrium with a heat reservoir, which is discussed next.

\subsection{HFT for pure states}
\label{sec:uHFT}

For any stationary normalized eigenvector $| \psi_n (\lambda) \rangle$ 
with the corresponding eigenvalue $E_n(\lambda)$ 
of a Hamiltonian $H(\lambda)$, i.e. $\langle \psi_n | \psi_n \rangle = 1$ 
and $E_n = \langle \psi_n | H | \psi_n \rangle$, the usual HFT states that 
\begin{equation}
\label{eqn:uHFT}
\frac{\partial E_n}{\partial \lambda} = \left\langle \psi_n \left| \frac{\partial H}{\partial \lambda} \right| \psi_n \right\rangle,
\end{equation}
where $\lambda$ is an arbitrary real parameter, explicitly appearing in $H$. 
This well-known theorem has proved to be very useful in many fields ranging 
from quantum chemistry, quantum statistics and many-body physics to molecular 
physics, with a lot of applications. 

In particular, the HFT has recently been used to derive a number of exact relations 
for strongly-correlated systems with short-range interations, in the context 
of ultra-cold quantum gases~\cite{braaten1, braaten2, werner1}. 
Since this theorem applies only for pure states, these exact relations 
strictly hold at zero temperature for any change or equivalently at finite 
temperatures for an adiabatic change. In order to generalize these relations to 
finite temperatures, here we derive the IHFT for a canonical ensemble of 
a fixed number of identical particles that are in thermal equilibrium 
with a heat reservoir.

\subsection{IHFT for mixed states in a canonical ensemble}
\label{sec:cHFT}

For this purpose, we assume the temperature $T$ is fixed, and the Hamiltonian 
$H$ is a function of some arbitrary real parameter $\lambda$. There are 
two complimentary ways to obtain the isothermal HFT in a canonical ensemble. 

The simplest way follows from the definition of the statistical Helmholtz 
Free energy $F$, i.e. $F = -T \ln Z$ in a unit of $k_B = 1$ here and 
throughout, where 
$
Z = \mathrm{Tr}(e^{-H/T})
$
is the canonical partition function of the system. Then, it is easy to show 
that (see also~\cite{son2, thomas2})
\begin{equation}
\label{eqn:FHFT}
 \frac{\partial F}{\partial \lambda} = \left\langle \frac{\partial H}{\partial \lambda} \right\rangle,
\end{equation}
where the ensemble average of any operator is defined by
$
\langle A \rangle = \mathrm{Tr}(\rho A)
$
and
$
\rho = e^{-H/T}/Z
$
is the density operator in the canonical ensemble. We recall that the 
derivatives with respect to $\lambda$ are evaluated at fixed $T$ here and throughout.
In addition, using the definitions of the thermodynamic Free energy, average 
energy and entropy, i.e. $F = E - TS$, $E = \langle H \rangle$ and 
$S = -\mathrm{Tr}(\rho \ln \rho) = -\partial F/ \partial T$, respectively, Eq.~(\ref{eqn:FHFT}) can be written as
\begin{equation}
\label{eqn:cHFT}
\frac{\partial E}{\partial \lambda} = \left( 1 - T\frac{\partial}{\partial T} \right) 
\left\langle \frac{\partial H}{\partial \lambda} \right\rangle.
\end{equation}
This is the isothermal HFT for mixed states in a canonical ensemble, 
and it will play a central role in the remaining parts of this manuscript. 

An alternative way of deriving Eq.~(\ref{eqn:cHFT}) is as follows. We start 
directly from the definition of $E$, i.e. $E = \mathrm{Tr}(\rho H)$, 
and evaluate the derivative to obtain
\begin{align}
\label{eqn:gHFT}
\frac{\partial E}{\partial \lambda} = 
\left\langle \left( 1 + \frac{E-H}{T} \right) \frac{\partial H}{\partial \lambda} \right\rangle.
\end{align}
In the intermediate steps, we used
$
\mathrm{Tr}(\partial e^{-H/T} / \partial \lambda) = -(1/T) \mathrm{Tr}(e^{-H/T} \partial H/ \partial \lambda)
$
and
$
\mathrm{Tr}(H \partial e^{-H/T} / \partial \lambda) = -(1/T) \mathrm{Tr}(H e^{-H/T} \partial H/ \partial \lambda),
$
both of which can be derived by writing
$
\mathrm{Tr}(A) = \sum_n \langle \psi_n | A | \psi_n \rangle
$
where $| \psi_n \rangle$ is the normalized eigenvector and $E_n$ is the corresponding 
eigenvalue of $H$. In addition, we used
$
\partial e^{-E_n/T} / \partial \lambda = \langle\psi_n| \partial e^{-H/T} / \partial \lambda |\psi_n\rangle,
$
which can be obtained from the usual HFT given in Eq.~(\ref{eqn:uHFT}).
For fixed $T$, starting from the definition
$
\langle A \rangle = \mathrm{Tr}(\rho A),
$
it is easy to show for the observables that commute with $H$, i.e. $[H,A] = 0$, that
$
\partial \langle A \rangle/\partial T = \langle (H - E) A \rangle/T^2.
$ 
Therefore, Eqs.~(\ref{eqn:cHFT}) and~(\ref{eqn:gHFT}) are equivalent.

Having derived the IHFT for a canonical ensemble, next we use this theorem
to obtain a number of exact relations for strongly-correlated 
ultra-cold quantum gases, which is the main purpose of this manuscript.

\section{Isothermal sweep theorems}
\label{sec:apps}

In this section, we use the IHFT given in Eq.~(\ref{eqn:cHFT}) to derive 
the isothermal sweep theorems (IST) for the free, average and trapping 
energies, and for the entropy, specific heat, pressure and atomic 
compressibility in a canonical ensemble.

\subsection{IST for the free, average and trapping energies}
\label{sec:energy}

As a first application of the IHFT, we consider the Hamiltonian that describes 
two-component Fermi gases in an external potential $U_\sigma(\mathbf{r})$, i.e.
\begin{align}
\label{eqn:fermihamiltonian}
 H = &\sum_\sigma \int d^3\mathbf{r} \psi_\sigma^\dagger(\mathbf{r}) \left[-\frac{\hbar^2 \nabla^2}{2m_\sigma} 
+ U_\sigma(\mathbf{r}) \right] \psi_\sigma(\mathbf{r}) \nonumber \\
&- g \int d^3\mathbf{r} \psi_\uparrow^\dagger(\mathbf{r}) \psi_\downarrow^\dagger(\mathbf{r}) 
\psi_\downarrow(\mathbf{r}) \psi_\uparrow(\mathbf{r}),
\end{align}
where $\psi_\sigma^\dagger(\mathbf{r})$ creates a pseudospin-$\sigma$ fermion 
with mass $m_\sigma$, and $g \ge 0$ is the strength of the short-range interaction.
In the following discussions, the numbers of $\uparrow$ and $\downarrow$ 
fermions need not be equal.
As usual the theoretical parameter $g$ can be written in terms of the 
experimentally more relevant scattering length $a$ via
$
1/g = -MV/(4\pi \hbar^2 a) + (M/\hbar^2) \sum_{k} (1/k^2),
$
where
$
M = 2m_\uparrow m_\downarrow / (m_\uparrow + m_\downarrow)
$
is twice the reduced mass of $\uparrow$ and $\downarrow$ fermions, 
and $V$ is the volume. This equation gives
$
g = -4\pi^2 \hbar^2 a / (M V \pi - 2 M V a k_c),
$
where $k_c$ is the momentum-space cut-off used in order to evaluate 
the ultraviolet divergent sum over $\mathbf{k}$.

Following the recent work~\cite{braaten1}, the contact parameter 
$C$ can be defined as
\begin{align}
\left\langle \frac{\partial H}{\partial a} \right\rangle &= \frac{M V g^2}{4\pi \hbar^2 a^2} 
\int d^3\mathbf{r} \langle \psi_\uparrow^\dagger(\mathbf{r}) \psi_\downarrow^\dagger(\mathbf{r}) 
\psi_\downarrow(\mathbf{r}) \psi_\uparrow(\mathbf{r}) \rangle, \\
\label{eqn:contact}
& = \frac{\hbar^2 C}{4\pi M a^2},
\end{align}
where we used $dg/da = -M V g^2/(4\pi \hbar^2 a^2)$. Note that $C$ 
is an extensive quantity that depends on both $a$ and $T$. 
This definition guarantees~\cite{braaten1} that the average 
energy $E$ of the system is of the desired form~\cite{tan1, combescot},
\begin{equation}
 E - \sum_\sigma \langle U_\sigma \rangle 
  = \sum_{\sigma,\mathbf{k}} \frac{\hbar^2 k^2}{2m_\sigma} \left[n_\sigma(\mathbf{k}) - \frac{C}{k^4} \right] 
    + \frac{\hbar^2 C}{4\pi M a},
\end{equation}
where $n_\sigma(\mathbf{k})$ is the momentum distribution of $\sigma$ fermions. 
As emphasized in~\cite{tan1}, this relation holds for any finite-energy state 
of the system, it does not matter whether the system is 
few- or many-body, superfluid or normal, weakly or strongly interacting, 
in equilibrium or nonequilibrium, at zero or finite temperature. 
Combining Eq.~(\ref{eqn:contact}) with the IHFT given in Eqs.~(\ref{eqn:FHFT}) 
and~(\ref{eqn:cHFT}), we obtain 
\begin{align}
\label{eqn:fenergy}
 \frac{\partial F}{\partial a} &= \frac{\hbar^2 C}{4\pi M a^2}, \\
\label{eqn:energy}
 \frac{\partial E}{\partial a} &= \frac{\hbar^2}{4\pi M a^2} 
\left(C - T\frac{\partial C}{\partial T}\right).
\end{align}
Here, the derivatives with respect to $a$ are evaluated at fixed $T$, and
therefore, Eqs.~(\ref{eqn:fenergy}) and~(\ref{eqn:energy}) correspond 
to the IST for the free and average energies, respectively.
Compared to the zero temperature expression for any change or equivalently 
the finite temperature expression for an adiabatic change, i.e. the adiabatic
sweep theorem for the energy~\cite{review, tan2}, the main difference 
in Eq.~(\ref{eqn:energy}) is an extra $T\partial C/\partial T$ term.
Similarly, we can also calculate
$
\partial F / \partial k_c = \hbar^2 C/(2\pi M)
$
and
$
\partial E / \partial k_c = [\hbar^2/(2\pi M)] (C - T \partial C / \partial T),
$
where we used $dg/dk_c = -M V g^2/(2\pi^2 \hbar^2)$.

For homogenous systems (no external potential), the $T$ dependence of $C$ 
has been recently calculated in the low, intermediate and 
high $T$ regimes as~\cite{yu}
\begin{eqnarray}
\label{eqn:lTC}
C &=& C_0 + \alpha_1 T^4, \,\,\, (T \ll T_c < T_F) \\
\label{eqn:lTC2}
C &=& C_0 + \alpha_2 T^2, \,\,\, (T_c < T \ll T_F) \\
\label{eqn:hTC}
C &=& \frac{\alpha_3}{T}, \,\,\, (T \gg \{T_F,T_a\}) 
\end{eqnarray}
where, up to the leading orders in $a$, $C_0 = 4\pi^2 n^2 a^2$,
$
\alpha_1 = 9\sqrt{3} \pi^6 n^2 a^2/(40 T_F^4)
$ 
and
$
\alpha_2 = -4(7\ln2-1)\pi^3 k_F n^2 a^3/(5 T_F^2)
$
in the BCS limit; $C_0 = 2\pi n/a$ and 
$
\alpha_1 = 2\pi^6 \sqrt{\pi k_F/a}(\partial a_m/\partial a)n/T_F^4
$
in the BEC limit; and $\alpha_3 = 8\pi^2 n^2/M$ for all couplings.
Here, $k_F$ is the Fermi momentum, $T_c$ is the critical temperature
for superfluidity, $T_F = \hbar^2 k_F^2/(2M)$ is the 
Fermi temperature, $T_a = \hbar^2/(Ma^2)$ is the binding energy of two fermions, 
$n = N/V = k_F^3/(3\pi^2)$ is the density of fermions, $a_m$ is the dimer-dimer 
scattering length between Cooper molecules 
(e.g. $a_m = 0.6a$ when $m_\uparrow = m_\downarrow$),
and the weakly-interacting BCS and BEC limits are characterized 
by $k_Fa \to 0^-$ and $k_Fa \to 0^+$, respectively. 
Note that our definition of $C$ is larger by a
factor of $4\pi^2$ compared to the definition of Ref.~\cite{yu}. 
Note also that the expression for intermediate $T$ regime, 
i.e. Eq~(\ref{eqn:lTC2}), is valid only in the BCS limit, 
and that $C$ has a maximum at a particular $T$, since $C_0$, $\alpha_1$, 
$\alpha_2$ and $\alpha_3$ are all positive constants with respect to $T$.
Using Eqs.~(\ref{eqn:lTC})-(\ref{eqn:hTC}) in Eq.~(\ref{eqn:energy}), we find
$
\partial E / \partial a = \hbar^2 (C_0 - 3\alpha_1 T^4)/(4\pi M a^2),
$
$
\partial E / \partial a = \hbar^2 (C_0 - \alpha_2 T^2)/(4\pi M a^2)
$
and
$
\partial E / \partial a = \hbar^2 C/(2\pi M a^2)
$
for the low, intermediate and high $T$ regimes, respectively.

Note also that taking the derivative of the Virial theorem given 
in Eq.~(\ref{eqn:uvirialf}) with respect to $a$, and using the IST for 
the energy given in Eq.~(\ref{eqn:energy}), we obtain
\begin{equation}
\label{eqn:Etrap}
 \frac{\partial E_{tr}}{\partial a} = \frac{\hbar^2}{16\pi Ma^2} 
\left( C + a\frac{\partial C}{\partial a} - 2T \frac{\partial C}{\partial T}\right),
\end{equation}
which corresponds to the IST for the trapping energy. Here,
\begin{equation}
\label{eqn:Etr}
E_{tr} = \frac{1}{2}\left\langle U + \frac{1}{2}\sum_{i = 1}^N \mathbf{r_i} \cdot \mathbf{\nabla_{r_i}}U \right\rangle
\end{equation}
is the effective trapping energy, which reduces to the trapping energy 
$\langle U \rangle$ in the case of harmonic trapping potentials.

\subsection{IST for the entropy and specific heat}
\label{sec:thermo}

The IHFT can also be used to find other thermodynamic relations. For instance,
we obtain an entropy relation by taking the derivative of $S = (E + F)/T$
with respect to $\lambda$, and using Eqs.~(\ref{eqn:FHFT}) and~(\ref{eqn:cHFT}),
leading to
\begin{equation}
\label{eqn:gentropy}
\frac{\partial S} {\partial \lambda} = 
-\frac{\partial}{\partial T} \left\langle \frac{\partial H}{\partial \lambda} \right\rangle.
\end{equation}
In addition, we obtain a specific heat relation (at constant volume) from 
$
C_V = \partial E / \partial T = T \partial S / \partial T,
$
leading to
\begin{equation}
\label{eqn:gsheat}
\frac{\partial C_V} {\partial \lambda} = 
- T \frac{\partial^2}{\partial T^2} \left\langle \frac{\partial H}{\partial \lambda} \right\rangle.
\end{equation}
In Eqs.~(\ref{eqn:gentropy}) and~(\ref{eqn:gsheat}) the derivatives with respect 
to $\lambda$ are evaluated at fixed $T$. We hope that these relations could be 
tested with thermodynamic measurements in strongly-interacting Fermi gases~\cite{thomas1}.

For ultra-cold quantum gases described by the Hamiltonian given in 
Eq.~(\ref{eqn:fermihamiltonian}), we can use Eq.~(\ref{eqn:contact})
in Eqs.~(\ref{eqn:gentropy}) and~(\ref{eqn:gsheat}), leading to
\begin{eqnarray}
\label{eqn:entropy}
\frac{\partial S} {\partial a} &=& -\frac{\hbar^2}{4\pi M a^2} \frac{\partial C}{\partial T}, \\
\label{eqn:sheat}
\frac{\partial C_V} {\partial a} &=& -\frac{\hbar^2 T}{4\pi M a^2} \frac{\partial^2 C}{\partial T^2}.
\end{eqnarray}
Here, the derivatives with respect to $a$ are evaluated at fixed $T$,
and therefore, Eqs.~(\ref{eqn:entropy}) and~(\ref{eqn:sheat}) correspond to
the IST for the entropy and specific heat, respectively. 
Equation~(\ref{eqn:entropy}) was first derived in Ref.~\cite{yu}.
Using Eqs.~(\ref{eqn:lTC})-(\ref{eqn:hTC}) in Eqs.~(\ref{eqn:entropy}) 
and~(\ref{eqn:sheat}), we find
$
\partial S / \partial a = -\alpha_1 T^3/(\pi M a^2)
$
and
$
\partial C_V / \partial a = -3\alpha_1 T^3/(\pi M a^2)
$
for the low $T$,
$
\partial S / \partial a = -\alpha_2 T/(2\pi M a^2)
$
and
$
\partial C_V / \partial a = -\alpha_2 T/(2\pi M a^2)
$
for the intermediate $T$, and
$
\partial S / \partial a = C /(4\pi M a^2 T)
$
and
$
\partial C_V / \partial a = -C /(2\pi M a^2 T)
$
for the high $T$ regimes. Therefore, since $\alpha_1$ and $\alpha_2$ 
are positive constants, while $\partial S / \partial a$ is negative for
low $T$, it becomes positive at high $T$, indicating that 
$\partial S / \partial a$ vanishes at a particular temperature above $T_F$.
Note also that $\partial S / \partial a \propto \partial C_V / \partial a$ 
in these regimes, and that the $T$ dependence of $C$ could be 
determined by the variation of either $S$ or $C_V$ with respect to $a$.

\subsection{IST for the pressure}
\label{sec:pressure}

For the Hamiltonian given in Eq.~(\ref{eqn:fermihamiltonian}), but in the absence
of the potential term, i.e. a homogenous system where $U(\mathbf{r}) = 0$, 
the IHFT can be used to derive the IST for the pressure.

In general, via a dimensional analysis, the Free energy $F$ can be written as
$
F(\eta_1,\dots,\eta_r) = (\hbar^2 \lambda^2/M) f(\lambda \eta_1, \dots, \lambda \eta_r)
$
where $\lambda$ has the dimension of the inverse of a length, $f$ is a dimensionless 
function of its parameters, and $\eta_j$ labels $r$ parameters 
(all with the dimension of a length) that $F$ 
may depend on for a given $H$. For homogenous ultra-cold quantum gases, since $F$ is a 
function of $T$, $V$, $N_\sigma$ and $a$, dimensional analysis~\cite{braaten2} 
requires that $F$ must satisfy
\begin{align}
\label{eqn:dimensional}
 2F &= - \sum_{q = 1}^r \eta_q \frac{\partial F}{\partial \eta_q}
&= -a\frac{\partial F}{\partial a} + 2T\frac{\partial F}{\partial T} - 3V\frac{\partial F}{\partial V}.
\end{align}
Using $F = E - TS$, $S = - \partial F / \partial T$, $P = - \partial F / \partial V$,
and the IST for the free energy given in Eq.~(\ref{eqn:fenergy}), we obtain
\begin{equation}
\label{eqn:pressure}
 P = \frac{2E}{3V} + \frac{\hbar^2 C}{12 \pi M a V}.
\end{equation}
Therefore, the universal pressure relation in a canonical ensemble 
is of the same form as the zero temperature one~\cite{review, tan2}.

Taking the derivative of Eq.~(\ref{eqn:pressure}) with respect to $a$, 
and using the IST for the energy given in Eq.~(\ref{eqn:energy}), we obtain
\begin{equation}
\label{eqn:ISTpressure}
 \frac{\partial P}{\partial a} = \frac{\hbar^2}{12\pi Ma^2 V} 
\left( C + a\frac{\partial C}{\partial a} - 2T \frac{\partial C}{\partial T}\right),
\end{equation}
which corresponds to the IST for the pressure. 
We recall that Eq.~(\ref{eqn:ISTpressure}) is derived for a homogenous 
system, and therefore, it is incorrect to compare it with 
Eq.~(\ref{eqn:Etrap}), and conclude that
$
3V \partial P / \partial a = 4 \partial E_{tr} / \partial a
$
for a trapped system.
Using Eqs.~(\ref{eqn:lTC})-(\ref{eqn:hTC}) in Eq.~(\ref{eqn:ISTpressure}), 
we find
$
\partial P / \partial a = \hbar^2 (3C_0 - 5\alpha_1 T^4) / (12\pi Ma^2 V)
$
in the BCS and
$
\partial P / \partial a = -5\hbar^2 \alpha_1 T^4 / (8\pi Ma^2 V)
$
in the BEC limit for the low $T$ regime,
$
\partial P / \partial a = \hbar^2 C_0 / (4\pi Ma^2 V)
$
in the BCS limit for the intermediate $T$ regime, and
$
\partial P / \partial a = \hbar^2 C/(4\pi Ma^2 V)
$
for the high $T$ regime. 
Therefore, since $\alpha_1$ is a positive constant, 
while $\partial P / \partial a$ is positive in the BCS limit, 
it becomes negative in the BEC limit in the low $T$ regime, 
indicating that $\partial P / \partial a$ vanishes at a 
particular scattering length around unitarity.

\subsection{IST for the atomic compressibility}
\label{sec:ic}

The isothermal atomic compressibility is defined as
$
\kappa_T = -V^{-1} \partial V/\partial P,
$
and it can be obtained by taking the derivative of Eq.~(\ref{eqn:pressure}) 
with respect to $V$ at constant $T$. 
Using Eq.~(\ref{eqn:fenergy}) to relate $\partial C / \partial V$ to $P$, 
$P = - \partial F / \partial V$, and $E = F + TS$ together with 
the Maxwell relation 
$
\left. \partial S / \partial V \right|_T = \left. \partial P / \partial T \right|_V,
$
we obtain
\begin{equation}
\label{eqn:kappa}
\frac{3}{\kappa_T} = 5P + a\frac{\partial P}{\partial a} - 2T \frac{\partial P}{\partial T}.
\end{equation}
This equation also follows from a dimensional analysis of the pressure $P$
which is a function of $T$, $V$, $N_\sigma$ and $a$, 
similar to the analysis that lead to Eq.~(\ref{eqn:dimensional}).
The $\partial P / \partial a$ term is given by Eq.~(\ref{eqn:ISTpressure}), 
and the $\partial P / \partial T$ term can be obtained by taking 
the derivative of Eq.~(\ref{eqn:pressure}) with respect to $T$, leading to
\begin{equation}
\label{eqn:sheat2}
C_V = \frac{3V}{2} \frac{\partial P}{\partial T} 
- \frac{\hbar^2}{8\pi Ma^2} \frac{\partial C}{\partial T},
\end{equation}
where $C_V = \partial E / \partial T$ is the specific heat at constant $V$.
Note that this equation relates the specific heat to the pressure and 
contact parameter. Using Eqs.~(\ref{eqn:ISTpressure}) and~(\ref{eqn:sheat}) 
in Eq.~(\ref{eqn:kappa}), we obtain
\begin{equation}
\label{eqn:kappa2}
 \frac{3}{\kappa_T} = 5P - \frac{4T}{3V}C_V + \frac{\hbar^2}{12\pi Ma V} 
\left( C + a\frac{\partial C}{\partial a} - 4T \frac{\partial C}{\partial T}\right),
\end{equation}
which relates the isothermal atomic compressibility to the pressure, specific heat
and contact parameter. When $C = 0$, Eq.~(\ref{eqn:kappa2}) is satisfied
for the ideal Fermi gases. The IST for the atomic compressibility can be
easily obtained by taking the derivative of Eq.~(\ref{eqn:kappa2})
with respect to $a$, and using Eqs.~(\ref{eqn:ISTpressure}) and~(\ref{eqn:sheat}).

Note that since the compressibility is related to the density 
fluctuations via the fluctuation-dissipation theorem,
$
\kappa_T = VT^{-1} (\langle \hat{N}^2 \rangle 
- \langle \hat{N} \rangle^2)/\langle \hat{N} \rangle^2,
$
where $\hat{N}$ is the density operator, it can be used to extract 
some thermodynamic information in atomic systems. Although this was
proposed as early as in 2005~\cite{iskin}, it has recently been possible
to extract this information for two-component Fermi gases, by measuring 
the density fluctuations and atomic compressibility~\cite{sanner, muller}. 
Combining Eq.~(\ref{eqn:kappa2}) with the fluctuation-dissipation
theorem, provides yet another universal relation that can be verified
with atomic systems.
Note also that the isoentropic (or adiabatic) compressibility $\kappa_S$
is related to the $\kappa_T$ via 
$
\kappa_S/\kappa_T = C_V/C_P,
$
where $C_P$ is the specific heat at constant $P$. Since the
specific heats are also related to each other via
$
C_P = C_V + T V \kappa_T (\partial P/ \partial T)^2,
$
using Eq.~(\ref{eqn:sheat2}), we obtain 
\begin{equation}
\label{eqn:kappaS}
\frac{1}{\kappa_S} = \frac{1}{\kappa_T} 
+ \frac{TV}{C_V}\left(\frac{2C_V}{3V} + \frac{\hbar^2}{12\pi Ma V} \frac{\partial C}{\partial T}\right)^2,
\end{equation}
which relates $\kappa_S$ to $\kappa_T$, $C_V$ and $C$. 
In atomic systems, it is easier to measure $\kappa_S$ than $\kappa_T$, and
Eq.~(\ref{eqn:kappaS}) can be used to extract the temperarure dependence 
of $C$, given that $\kappa_T$ can be extracted from the fluctuation-dissipation 
theorem~\cite{sanner, muller}.

Having derived the IST for the free, average and trapping energies, and 
for the entropy, specific heat, pressure and atomic compressibility 
in a canonical ensemble, next we derive the Virial theorem.

\section{Virial theorem for trapped systems}
\label{sec:virial}

In this section, we use the IHFT given in Eq.~(\ref{eqn:cHFT}) to derive 
the Virial theorem~\cite{son2, thomas1, tan3, werner1} in a canonical
ensemble. This can be most easily achieved following the recent 
work on the zero-temperature case~\cite{werner1}.

\subsection{Virial theorem in a canonical ensemble}
\label{sec:gvirial}

For this purpose, consider a general Hamiltonian 
$
H = K + I + U,
$
that describes $N$ particles with arbitrary statistics in arbitrary dimensions, 
where $K$ is the kinetic energy, $I$ is the interaction, 
and $U$ is an arbitrary external potential. 
For ultra-cold quantum gases, the external potential is simply
$U = \sum_{i = 1}^N U_i(\mathbf{r_i})$, where $U_i(\mathbf{r_i})$ has 
approximately harmonic dependence on the position $\mathbf{r_i}$ of the particles. 

In general, via a dimensional analysis, $U$ can be written as
$
U(\mathbf{r_1}, \dots, \mathbf{r_N}) = (\hbar^2 \lambda^2/M) u(\lambda \mathbf{r_1}, \dots, \lambda \mathbf{r_N}),
$
where $\lambda$ has the dimension of the inverse of a length, 
and $u$ is a dimensionless function of its parameters. Therefore, we can use the 
IHFT given in Eq.~(\ref{eqn:cHFT}) to obtain 
$
\lambda \partial E / \partial \lambda = 4( 1 - T \partial / \partial T) E_{tr},
$
where $E_{tr}$ is the effective trapping energy defined in Eq.~(\ref{eqn:Etr}).
In addition, we can write the energy $E$, via again a dimensional analysis, as
$
E(\ell_1,\dots,\ell_p) = (\hbar^2 \lambda^2/M) e(\lambda \ell_1, \dots, \lambda \ell_p),
$
where $\ell_q$ labels $p$ parameters (all with the dimension of a length) that $E$ 
may depend on for a given $H$, and $e$ is a dimensionless function of its parameters. 
Evaluating the derivative with respect to $\lambda$ for fixed values of $\ell_q$, 
we obtain
$
\lambda \partial E / \partial \lambda = 2E + \sum_{q = 1}^p \ell_q \partial E / \partial \ell_q.
$
Then, the Virial theorem is obtained by combining these two analysis, leading to
\begin{equation}
\label{eqn:virial}
 E = 2\left( 1 - T\frac{\partial}{\partial T} \right) E_{tr}
- \frac{1}{2}\sum_{q = 1}^p \ell_q \frac{\partial E}{\partial \ell_q}.
\end{equation}
Compared to the zero temperature expression~\cite{werner1}, the main differences 
here are an extra $T \partial/\partial T$ term in front of the potential terms, 
and an extra $\ell_q$ term associated with the temperature.

\subsection{Trapped quantum gases}
\label{sec:uvirial}

In particular, for the ultra-cold quantum gases, which are well described by 
the s-wave scattering length $a$, Eq.~(\ref{eqn:virial}) reduces to
$
(1 - T\partial / \partial T) (E - 2E_{tr}) = -(a/2) \partial E / \partial a.
$
Note that for finite $k_c$, i.e. nonzero interaction range, there would be 
an additional 
$
-(k_c/2) \partial E / \partial k_c = -\hbar^2 C k_c /(4 \pi^2 M)
$
term on the right hand side of this equation. Furthermore, using the 
IST for the average energy given in Eq.~(\ref{eqn:energy}) for the 
last term, solution of the resultant differential equation~\cite{notevirial} 
can be written as
\begin{equation}
\label{eqn:uvirial}
E = 2E_{tr} - \frac{\hbar^2 C}{8\pi M a} + \kappa T,
\end{equation}
where $\kappa$ is a real constant independent of $T$. 
This is the most general form of the Virial theorem in a canonical ensemble. 
Compared to the zero temperature expression~\cite{tan3, werner1, braaten1}, 
the main difference here is an extra $\kappa T$ term. In the unitarity 
$a \to \pm \infty$ limit, it was shown via a dimensional analysis that 
the Virial theorem in a canonical ensemble does not have the last 
term~\cite{thomas1}, and hence, we know that $\kappa$ vanishes in this limit, 
i.e. $\kappa_{a \to \pm \infty} = 0$. We suspect $\kappa = 0$ for all $a$, 
however, since a nonzero $\kappa$ is allowed in general, this possibility 
needs to be clarified via other means.

Similar to the dimensional analysis that lead to Eq.~(\ref{eqn:dimensional}),
it can be shown that Eq.~(\ref{eqn:uvirial}) follows from a dimensional 
analysis of the average energy $E$ supplied with Eq.~(\ref{eqn:energy}). 
Note that $E$ is a function of $T$, $N_\sigma$, $a$ and the trapping 
frequency $\omega$ for a trapped system. However, applying a dimensional 
analysis to the free energy $F$ of trapped systems, which is also a function 
of $T$, $N_\sigma$, $a$ and $\omega$, and using Eq.~(\ref{eqn:fenergy}), 
we obtain Eq.~(\ref{eqn:uvirial}) with $\kappa = 0$~\cite{braatennote}. 
Therefore, we conclude that $\kappa = 0$ for all parameter space, i.e.
\begin{equation}
\label{eqn:uvirialf}
E = 2E_{tr} - \frac{\hbar^2 C}{8\pi M a},
\end{equation}
and that the Virial theorem in a canonical ensemble is of the same form 
as the zero temperature one~\cite{tan3, werner1, braaten1}. 
We recall that taking the derivative of Eq.~(\ref{eqn:uvirialf}) 
with respect to $a$, and using the IST for the energy given in 
Eq.~(\ref{eqn:energy}), we obtained the IST for the trapping energy 
given in Eq.~(\ref{eqn:Etrap}).

\section{Conclusions}
\label{sec:conclusions}

To conclude, first we derived the isothermal Hellmann-Feynman theorem 
that is suitable for mixed states in a canonical ensemble. 
Then, we obtained the isothermal magnetic-field sweep theorems for 
the free, average and trapping energies, and for the entropy, 
specific heat, pressure and atomic compressibility of 
strongly-correlated ultra-cold quantum gases. 
We applied the sweep theorems to two-component Fermi gases in the 
weakly interacting BCS and BEC limits, and showed that the 
temperature dependence of the contact parameter could be determined 
by the variation of either the entropy or specific heat 
with respect to the scattering length. 
We also obtained the Virial theorem in a canonical ensemble, 
and discussed its implications for quantum gases. 

One of the major challenges for the experiments with ultra-cold quantum 
gases is the lack of a precise thermometry, and even 
the measure of the temperature itself for strongly-interacting
Fermi gases is a challenging problem~\cite{thermo}. 
On one hand, this makes it more difficult to perform an isothermal 
(constant temperature) magnetic-field sweep at ultra-cold 
temperatures, compared to an adiabatic (constant entropy) 
sweep that is routinely performed in atomic systems, while tuning 
the scattering length. On the other hand, it is theoretically more 
easier to calculate thermodynamic quantitites at constant temperature, 
e.g. the calculation of isothermal atomic compressibility is much
easier than the isoentropic atomic compressibility in the BCS-BEC 
crossover. Therefore, the isothermal Hellmann-Feynman and sweep 
theorems that we discussed in this paper are probably most useful 
for other theoretical or numerical studies, and possibly for
some special experiments where the thermometry is not an issue.

\section{Acknowledgments}
\label{sec:ack}

The author thanks Eric Braaten and Felix Werner for comments. This work 
is financially supported by the Marie Curie International Reintegration 
(FP7-PEOPLE-IRG-2010-268239) and Scientific and Technological Research 
Council of Turkey's Career (T\"{U}B$\dot{\mathrm{I}}$TAK-3501) Grants.


\begin{thebibliography}{99}
\bibitem{notevirial} After the substitution, the resultant equation is of the form 
$(1- T \partial / \partial T)y = 0$, for which the general solution can be written
as $y = \kappa T$ where $\kappa$ is a real constant independent of $T$.
\end{thebibliography}
\end{document}